\documentclass[twocolumn,showpacs,prl,amsmath,amssymb,superscriptaddress]{revtex4}
\usepackage{graphicx}
\usepackage{amsmath}
\usepackage{amssymb}
\usepackage{color}
\usepackage{verbatim}
\usepackage{hyperref}
\usepackage{epsf}
\usepackage{graphics}
\usepackage{natbib}
\usepackage{dcolumn}
\usepackage{bm}
\begin{document}

\title{Dimensional crossover of a frustrated distorted kagome Heisenberg Model: Application to FeCrAs}

\author{Travis E. Redpath}
\affiliation{University of Manitoba, Winnipeg, Manitoba, Canada, R3T 2N2}
\author{John M. Hopkinson}
\email{hopkinsonj@brandonu.ca}
\affiliation{Brandon University, Brandon, Manitoba, Canada, R7A 6A9}
\affiliation{University of Manitoba, Winnipeg, Manitoba, Canada, R3T 2N2}
\author{Alton A. Leibel}
\affiliation{Brandon University, Brandon, Manitoba, Canada, R7A 6A9}
\author{Hae-Young Kee}
\affiliation{University of Toronto, Toronto, Ontario, Canada, M5S 1A7}
\date{\today}

\begin{abstract}

Motivated by recent experimental work on FeCrAs, we study the magnetic properties of classical spin models on the hexagonal Fe$_2$P/ZrNiAl structure (space group $P{\bar{6}}2m$). When both transition metal/rare earth sites carry magnetic moments, one has alternating distorted kagome and triangular lattice layers. Each point of the triangular lattice lies at the centre of a distorted hexagon and actually corresponds to three isolated magnetic atoms: a ``trimer''. We show that a simple model consisting of antiferromagnetically correlated Heisenberg spins on the frustrated distorted kagome lattice coupling either ferromagnetically or antiferromagnetically to Heisenberg spins at the average position of a trimer, leads to a rich phase diagram, including the possible selection of a $(\frac{1}{3},\frac{1}{3},0)$ magnetic order as may have been seen in FeCrAs.

\end{abstract}
\maketitle

\section{Introduction}

Materials featuring magnetic moments situated on corner-shared triangles or tetrahedra
provide an exciting domain for the discovery of exotic physics.  Recently one class of such {\it{geometrically
frustrated magnets}} (GFM), spin-ice materials, garnered attention due to the possibility of interpreting the
excitations from the ground states as effectively deconfined magnetic monopoles {\cite{sondhimoessner}}.
Several experimental groups have claimed {\cite{experimentalmonopoles}} to have observed these
so-called monopoles, and have pointed out that they would likely pass several of the experimental tests for
the existence of magnetic monopoles. Of particular interest was that these fractional spin excitations appear to be
deconfined or essentially free in three dimensions.\cite{sondhimoessner}
Recently we have shown that spin ice physics should also arise on the three-dimensional trillium {\cite{us1}}
 and hyperkagome {\cite{us2}} lattices.

In a second class of GFM, Heisenberg spin systems, a long sought goal has been
the realization of a truly three-dimensional quantum spin liquid.  Indeed, the belief in the existence of such
a state strongly influenced the original resonating valence bond model of high temperature
superconductivity{\cite{Anderson87}}.  Therefore it was with great excitement that the community welcomed the
Mott insulating Na$_4$Ir$_3$O$_8$ which features antiferromagnetically correlated spin 1/2 magnetic moments and does
not magnetically order to very low temperatures{\cite{TakagiPRL2007}}.  These phenomena are well captured by a simple
classical Monte Carlo treatment of the Heisenberg model on the three dimensional corner-shared equilateral triangle hyperkagome
lattice{\cite{hopkinsonprl2007}}, which additionally predicted the onset of a finite temperature nematic transition.  Early quantum treatments of the physics of this material have argued that at least over a finite temperature range one has indeed found a quantum spin liquid{\cite{lawler1,zhou,lawler2,chen}} or at least a valence bond crystal.{\cite{bergholtz}}

 It has been considerably more difficult to experimentally realize such exotic physics on the simpler two-dimensional kagome net, perhaps because materials are inherently three-dimensional.   Indeed, after decades of searching, the first experimental realization of antiferromagnetic correlations on a structurally perfect spin 1/2 kagome lattice, ZnCu$_3$(OH)$_6$Cl$_2$,\cite{shores,helton} appeared only in 2005, followed soon after by the related ZnCu$_3$(OD)$_6$Cl$_2$ which shows evidence of a valence bond solid state { \cite{lee}}.  However, even in these materials questions remain about the realization of a purely kagome lattice, given the recently measured non-magnetic site disorder of the order of 6\% within the magnetic Cu plane\cite{devries}.

Interestingly, a large class of magnetic materials including some (non-superconducting) iron pnictides and the langasites form in a distorted kagome structure with a nearest neighbour topology equivalent to that of the kagome lattice.  Among these, Nd langasite (Nd$_3$Ga$_5$SiO$_{14}$) has previously been studied as a spin liquid candidate\cite{langasites}, while FeCrAs shows intriguing thermodynamic and transport signatures.\cite{wu}  Magnetically, FeCrAs has been seen\cite{swain} to exhibit a $\sqrt{3} \times \sqrt{3}$  magnetic ordering with wavevector of $(\frac{1}{3},\frac{1}{3},0)$ below 125K. 

While the nature of the spin-$\frac{1}{2}$ quantum Heisenberg model on the kagome lattice remains an important open problem\cite{diep}, 
the classical antiferromagnetic Heisenberg model has been shown\cite{chalker,huse,reimers,chubukov,harris} 
to have degenerate so-called $q=0$ and $\sqrt{3} \times \sqrt{3}$ ground states, and exhibit an order by disorder transition to select the latter at $T\rightarrow 0$, resulting in a local nematic ordering with a diverging correlation length in the limit $T\rightarrow 0$\cite{chalker}. {\it{Why then would FeCrAs choose to magnetically order at 125 K into a $(\frac{1}{3},\frac{1}{3},0)$ ground state?}}

 Could it be that nearest and next nearest neighbor exchange plus Dzyaloshinskyy-Moriya interaction terms are all necessary to a description of the physics on the distorted kagome lattice as has been studied by Gondek {\it{et al.}}\cite{gondek} for XY spins, which gives rise to interesting magnetic structures, including one with a $\left(\frac{1}{3},\frac{1}{3}\right)$ propagation vector?  The nearly isotropic resistivity and magnetic susceptibility, and the existence of the phase transition at finite temperatures speak to the possibility that we have a truly three dimensional material, naturally leading us to ask what the effect of a magnetic coupling between frustrated distorted kagome planes would lead to.

 Curiously, the addition of an interlayer coupling between stacked kagome planes to an antiferromagnetic Heisenberg model, and its role in leading to possible magnetic order, appears not to have received much attention.  
The spin-$\frac{1}{2}$ quantum antiferromagnetic Heisenberg model has been treated on the stacked kagome lattice by D. Schmalfu\ss\ {\it{et al.}}\cite{schmalfuss} using a spin-rotation-invariant Green's function method (supported by classical spin-wave calculations), finding the system to remain ``short-range ordered independent of the strength and sign of the interlayer coupling.''
 For classical spins, M. Zelli {\it{et al.}}\cite{zelli} used a short-time dynamics Monte Carlo method, scaling analysis and Binder cumulants to study the ordering properties of a generalized antiferromagnetic Heisenberg model on a stacked triangular lattice. A limiting case of this model is the classical Heisenberg model on the stacked kagome lattice for which they found no evidence of long range order.

 However, when the couplings between kagome planes are mediated by other magnetic sites,
a dramatically different result can be expected.
FeCrAs is an example of such coupled kagome systems in which 
both Fe and Cr are magnetic ions, and the dominant magnetic interaction is believed to be an antiferromagnetic Heisenberg
coupling between local moments on the Cr sites. 
The Cr sites form stacked distorted kagome planes. Three nearest Fe ions form a trimer (triangle), which can be viewed as an effectively single magnetic site. These effective Fe molecules then form triangular planes between Cr kagome planes. 
Therefore, FeCrAs can be viewed as alternating distorted kagome and triangular planes as shown in Fig. \ref{figure1}. A microscopic derivation of this physics, and the validity of our approximation will be presented elsewhere.\cite{jrau}

 In this paper, we propose a minimal microscopic model that captures the magnetic ordering
in FeCrAs at finite temperatures. Using MC simulation, we find that any small finite coupling between the Fe and Cr sites
leads to a coplanar $\sqrt{3} \times \sqrt{3}$ magnetic order consistent with existing neutron scattering data on FeCrAs.\cite{swain} As the coupling between Fe and Cr increases, there appear two finite temperature transitions. At high temperatures one has a transition to a ferri(ferro)magnetic state followed at low temperatures by a canting of each spin to realize a component of each spin which adopts a $\sqrt{3}\times \sqrt{3}$ order of some fraction of the spin.
%In addition to $\sqrt{3} \times \sqrt{3}$ coplanar ordering at low temperatures, spins on Cr and Fe cant away
%from the plane forming a ferrimagnet at high temperatures.
At larger interplane couplings still this gives way to ferri(ferro)magnetic order, as is commonly seen in other members of this family. 
This implies that we may have a large family of previously unrecognized frustrated magnets, and should have implications for uniaxial pressure experiments on FeCrAs, which can be  tested by a future experiments.

\section{general properties and lattice structure}

\subsection{Lattice structure}

Materials forming in the hexagonal Fe$_2$P/ZrNiAl structure with space group $P\bar{6}2m$\cite{P321} feature two site types potentially occupied by magnetic atoms labeled 3$f$ and 3$g$ as seen in Table \ref{TableI} and shown in Fig. \ref{figure1}. When in a given layer the value of $\frac{1}{3}<x_i<\frac{2}{3}$ ($i=1$ or $2$), the magnetic atoms belonging to that layer are arranged in a distorted kagome net, topologically equivalent to the kagome net for nearest neighbour couplings.  Thus antiferromagnetic interactions or ferromagnetic interactions with a strong single ion anisotropy would be strongly frustrated, although away from the special point $x_i=\frac{1}{2}$, which corresponds to the kagome lattice, the lattice lacks inversion symmetry so that Dzyaloshinkii-Moriya terms can arise.  Alternately, when $0<x_i<\frac{1}{3}$ or $\frac{2}{3}<x_i<1$, nearest neighbour bonds between magnetic sites form isolated triangles about the centers of the distorted hexagons in the layers directly above and below.  The special limit $x_i = \frac{1}{3}$ in a layer corresponds to the triangular lattice.  For simplicity in our model below we will take the limit $x_1 = 0$ which treats the isolated trimers as all being at a single point (0,0,0) of the hexagonal lattice.
 
\begin{table}[hbtp]
\begin{tabular}{|c|c|}
%\hline
\hline
Site type& Atomic position \\
\hline
\hline
3$f$&($\frac{x_1 a}{2}$,-$\frac{\sqrt{3}x_1 a}{2}$,0), ($\frac{x_1 a}{2}$,$\frac{\sqrt{3}x_1 a}{2}$,0), (-$x_1a$,0,0)\\
\hline 
3$g$&($\frac{x_2 a}{2}$,-$\frac{\sqrt{3}x_2 a}{2}$,$\frac{c}{2}$),($\frac{x_2 a}{2}$,$\frac{\sqrt{3}x_2 a}{2}$,$\frac{c}{2}$), (-$x_2 a$,0,$\frac{c}{2}$)\\
\hline
\hline
Name& Lattice vector\\
\hline
\hline

$\overrightarrow{A_1}$&($\frac{a}{2}$,-$\frac{\sqrt{3}a}{2}$,0)\\
\hline
$\overrightarrow{A_2}$&($\frac{a}{2}$,$\frac{\sqrt{3}a}{2}$,0)\\
\hline
$\overrightarrow{A_3}$&(0,0,$c$)\\
\hline

\end{tabular}
\caption{\label{TableI} Atomic positions of potentially magnetic atoms (3$f$,3$g$) in the Fe$_2$P/ZrNiAl structure, and hexagonal lattice vectors $\{{\bf{A}}_1,{\bf{A}}_2,{\bf{A}}_3\}$.  Generically $x_1<\frac{1}{3}$ which leads a layer of isolated ``trimers'', and $\frac{1}{3}<x_2<\frac{2}{3}$ which leads to a distorted kagome lattice of corner-shared equilateral triangles except at the special point $x_2=\frac{1}{2}$ which corresponds to a perfect kagome lattice.  
}
\end{table} 
\subsection{Relevant magnetic materials}

A large number of magnetic materials form in the hexagonal Fe$_2$P/ZrNiAl structure.  It has been estimated\cite{gondek} that almost 30\% of all rare-earth-transition metal compounds RTX crystallize in this form, with the rare earth metal generally sitting in a distorted kagome site.  Among these, antiferromagnetic order is seen in RAgSi for R=$\{$Dy,Ho,Er,Tb$\}$\cite{baran} and RAgGe for R=$\{$Gd,Er,Tb,Dy,Ho$\}$\cite{baran2}, which are metallic materials for which the RKKY interaction and strong crystalline electric fields have been thought to be relevant.  Antiferromagnetic Curie-Weiss offsets have likewise been seen in TbPtIn and TmAgGe, which have been argued to be describable in terms of a triple coplanar Ising model with large moments consistent with R$^{3+}$.\cite{morosan}  A large number of TT'X materials also form in this structure, some showing evidence of ferro- (or possibly ferri-) magnetic order, others showing evidence of antiferromagnetic order.  With T=Ni and X=P, T'=$\{$Mo,W,Fe,Co,Cr\cite{bacmann}$\}$ are known as are NiMnAs\cite{guerin} and MnTiP. Known arsenides feature TT' =$\{$CrPd,FeV,MnRu,CoCr$\}$\cite{guerin}.  Known\cite{guerin} silicides include TT'=$\{$TiMn,ZrRu,NbCr,NbMn$\}$, and known\cite{guerin} germanides include TT'=$\{$NbCr,HgFe,NbMn,TiCo$\}$.  Amongst the langasites one has at least two magnetic members\cite{iwataki}, Nd$_3$Ga$_5$SiO$_{14}$ and Pr$_3$Ga$_5$SiO$_{14}$, although in these materials the trimer lattice is occupied by Ga which is unlikely to have a magnetic moment.

\subsection{FeCrAs}

In FeCrAs it is believed{\cite{swain}} that there is perfect order of the Cr atoms in the 3$g$ sites, and the Fe atoms in the 3$f$ sites.  Swainson {\it{et al.}}\cite{swain} claim that for this material $x_1 =0.760(2)$, $x_2=0.436(2)$, $a=6.0676\AA$, and $c=3.6570\AA$, although these detailed values do not enter into our calculations of the heat capacity and magnetic susceptibility of the model presented below (which have additionally taken the $x_1\rightarrow 0$ (or equivalently 1) limit).  For these values, we note that the nearest neighbour Cr-Cr bonds form a distorted kagome lattice, while the Fe-Fe bond distance is quite short.

 FeCrAs is an unusual magnetically frustrated iron pnictide that has recently garnered interest as a possible novel non-Fermi liquid\cite{wu}. FeCrAs shows\cite{wu} Fermi liquid behaviour with a specific heat coefficient $\gamma$ of $30mJ/mol K^2$ which indicates that it may be metallic but this contrasts with a resistivity which increases almost isotropically from $800K$ to $80mK$\cite{wu}. Apart from a feature near a magnetic ordering transition in the a-axis resistivity, this increase at low temperatures is monotonic and below $10K$ appears to follow a weak power law $\delta\rho\sim -T^\alpha$ ($\alpha\sim0.6-0.7$) along both the a and c-axes\cite{wu} rather than the $\ln T$ one might expect from Kondo physics.

\begin{figure}[here]
\includegraphics[scale=0.21]{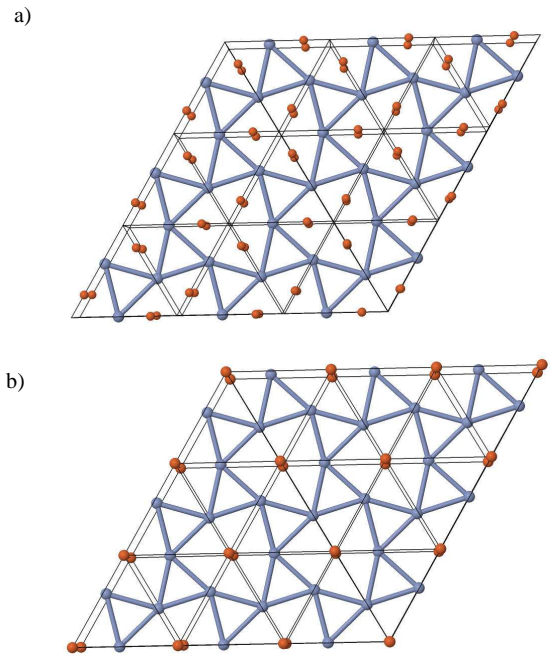}
\caption{(Color online) a) The FeCrAs lattice with (blue) circles at Cr (3g) sites joined to nearest neighbors and (red) circles at Fe (3f) sites. b) The groups of three Fe sites are taken to act together as a single Fe trimer.\label{figure1}}
\end{figure}

 At low temperature ($T \sim 125K$) FeCrAs shows a magnetic transition with an ordering wave vector of $\vec{k}=(\frac{1}{3},\frac{1}{3},0)$\cite{wu,swain}. From the susceptibility measurement, the magnetic interaction can be described as
an antiferromagnetic Heisenberg model between local moments on the Cr sites ($\mu\approx 1.22 \mu_B$\cite{swain}) that form distorted kagome planes as shown 
in Fig. \ref{figure1}(a). In between the Cr planes lie Fe atom trimers which have a small moment (if any $\mu\le(0.1\pm 0.03)\mu_B$\cite{swain}) and lie equidistant directly below and above the centre of the distorted hexagons of Cr in adjacent planes. Since according to the Mermin Wagner theorem one does not expect long range order to arise in a purely two dimensional system\cite{merminwagner}, we believe that at finite temperature an interplay between frustration and dimensionality may give rise to this behavior. Namely, we posit a nearest neighbour coupling between the distorted kagome Cr planes and the Fe trimers that lie between. A competing candidate for the origin of the dimensional crossover to allow a finite temperature phase transition if the Fe sites do not hold magnetic moments would be a dipolar interaction between neighboring Cr planes, although such a picture would need to account for the surprisingly high ordering temperature, given that dipolar interactions are estimated to be only of the order $1.27\times10^{-4}K$ here.

\section{Model}

 We consider a classical antiferromagnetic Heisenberg model between nearest neighbour Cr spins as has been previously 
studied in the context of the Nd langasites by Robert {\it{et al.}}\cite{langasites}.
For simplicity, to address the interesting dimensional crossover question, we assume that the iron atom trimers act as a single classical spin (see Fig. \ref{figure1}(b)) and that there is a Heisenberg coupling between each of these Fe trimers and Cr sites. The model is written:
\begin{equation}
H = J_1 \sum_{\langle ij \rangle}\vec{s}\ _i^{Cr}\cdot\vec{s}\ _j^{Cr}\ +\ J_2\sum_{\langle ik \rangle}\vec{s}\ _i^{Cr}\cdot\vec{s}\ _k^{Fe}\text{,}\label{equation1}
\end{equation}
where $J_1>0$ is an antiferromagnetic coupling and $J_2$ may be antiferromagnetic ($J_2>0$) or ferromagnetic ($J_2<0$). For this simple model, we assume that the spin coupling between the Cr and Fe sites is the same for the two inequivalent (but approximately equal) neighbor distances from the trimers. Thus each Cr site has 4 Fe neighbors with the same magnetic interaction in this model. Noting that the magnitude of any non-zero net spin on the Fe trimer can be (at least energetically) absorbed into the definition of $J_2$, we proceed to describe the general phase diagram as $\frac{J_2}{J_1}$ is varied from $0$ to $\inf$ keeping $J_1+|J_2|=1$ and unit spins on each site.

\subsection{Ground State Energy}

\begin{figure}
\includegraphics[scale=0.21]{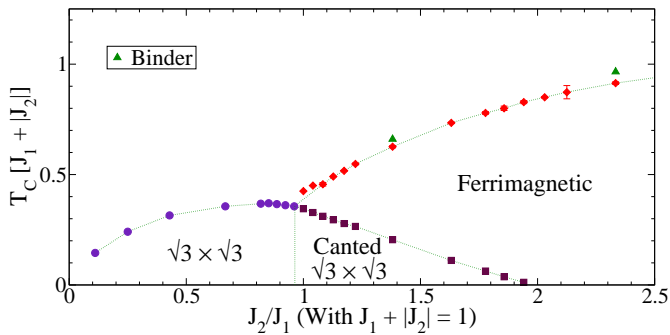}
\caption{(Color online) The phase diagram of FeCrAs as determined from peaks of heat capacity vs. temperature plots of Monte Carlo results at $L = 6$. Also included are the transition temperatures obtained from the Binder cumulant. These temperatures lie only slightly above the results obtained from the heat capacity data, indicating that the ferrimagnetic transition is not an artifact of the finite lattice size. The dotted line is a guide for the eyes. For $J_2 < 0$ the ferrimagnetic region is simply replaced by a ferromagnetic region.
\label{figure2}}
\end{figure}

The ground state energy for various values of $J_1$ and $J_2$ was found using numerical minimization. As we increase $J_2$ from zero, we see that initially (i.e. $\frac{J_2}{J_1} = 0$) we recover the known ground states of the Heisenberg model on the kagome lattice, which feature $120^\circ$ rotated spins on each of the corners of the equilateral triangles of the Cr sites, and a randomly oriented spin at the Fe site. At the opposite extreme, when $\frac{J_2}{J_1} \rightarrow \infty(-\infty)$, we have a ferrimagnetic(ferromagnetic) structure with Cr spins in the plane aligned antiparallel(parallel) to Fe spins, but parallel to each other. In between the spin structure has components of each structure, at each Cr site adopting a form,

\begin{equation}
\vec{S} = (\cos(\phi)\vec{s}_{120}, \sin(\phi))\text{.}\label{equation2}
\end{equation}

That is, for any three Cr spins forming a triangle, one can choose $\phi$ such that the spins are given by $(\cos(\phi), 0, \sin(\phi))$, $(-\frac{\cos(\phi)}{2}, \frac{\sqrt{3}\cos(\phi)}{2}, \sin(\phi))$, and $(-\frac{\cos(\phi)}{2},-\frac{\sqrt{3}\cos(\phi)}{2}, \sin(\phi))$. The Fe spins are then opposite(parallel) to the sums of their nearest two spins. Since the Fe spins lie above and below the centers of the distorted hexagons of the lattice, this implies that each hexagon has only two spin types around it. That is, the so-called $\sqrt{3}\times\!\sqrt{3}$ ordering of the kagome lattice is selected over the $q = 0$ state by the presence of the Fe atoms. This explains why there is a $(\frac{1}{3},\frac{1}{3},0)$ periodicity in the magnetic lattice as reported by Wu {\it{et al.}}\cite{wu}, as it takes three unit cells in each of the in-plane lattice vector directions for the spin structure to repeat. Next, let us compute the energy of this ground state.
For completeness, the directions of the Fe atoms can be taken as $\frac{-sgn(J_2)}{\sqrt{1+3\sin^2(\phi)}} ( \frac{\cos(\phi)}{2}, \frac{\sqrt{3}\cos(\phi)}{2}, 2\sin(\phi))$, $\frac{-sgn(J_2)}{\sqrt{1+3\sin^2(\phi)}}( \frac{\cos(\phi)}{2}, -\frac{\sqrt{3}\cos(\phi)}{2}, 2\sin(\phi))$, and $\frac{-sgn(J_2)}{\sqrt{1+3\sin^2(\phi)}}(-\cos(\phi), 0, 2\sin(\phi))$. Carrying out the dot products we find that the energy per spin (noting that there are 3 Cr and 1 Fe sites per unit cell in our model) is then given by,

\begin{equation}
\epsilon = 3\left(J_1(\frac{3}{2}\sin^2(\phi)-\frac{1}{2}) - |J_2|\sqrt{1+3\sin^2(\phi)}\right)\text{.}\label{equation3}
\end{equation}

Minimization of $\epsilon$ with respect to $\phi$ yields possible solutions of $\sin(\phi) = 0$, $\cos(\phi) = 0$ and $\frac{|J_2|}{J_1} = \sqrt{1+3 \sin^2(\phi)}$ (the latter being valid only for $1<\frac{|J_2|}{J_1}<2$). The former ($\phi = 0$), corresponds to the spins maintaining a purely $120^\circ$ rotated spin structure in the Cr plane, with the spins on the Fe sites coplanar, and opposite(parallel) to the spin sum of their nearest Cr sites. The energy of this state can be expressed as $-\frac{3}{2}J_1 - 3|J_2|$. This nicely reproduces our minimized energies until $\frac{|J_2|}{J_1} = 1$. At this point, the Cr spins begin to cant upwards uniformly from the $120^\circ$ rotated spin structure, while the Fe spins begin to tilt downward (upward if ferromagnetic). The energy of this spin structure is given by $-3J_1 - \frac{3J_2^2}{2J_1}$ , which minimizes the energy until $\frac{|J_2|}{J_1} = 2$, at which point $\phi = \frac{\pi}{2}$. Beyond this point, the spin structure is ferrimagnetic(ferromagnetic), with an energy per unit spin of $3J_1 - 6|J_2|$ which corresponds to all Cr spins parallel, and Fe spins antiparallel(parallel).

\section{Monte Carlo Simulation}

 Monte Carlo simulations were carried out using the Metropolis algorithm with periodic boundary conditions on a lattice with hexagonal unit cells with a total side length of $L$ unit cells containing $4L^3$ sites in total ($\frac{1}{4}$ being Fe trimers and $\frac{3}{4}$ being Cr atoms). The bulk of the analysis was done using a lattice size of $L=6$ which corresponds to $864$ spin sites. At each temperature $2\times 10^5$ Monte Carlo steps were used to equilibrate the system and a further 
$2 \times 10^5$ were used to calculate the averages of physical quantities, where one Monte Carlo step was taken to, on average, attempt one update per site. Error bars as reported correspond to the standard deviation of our averages over four independent trials.  Each update attempt has a randomly chosen magnitude, $\delta$, and direction relative to the initial spin direction. The value of $\delta$ was chosen so that around $50\%$ of the attempted spin updates are accepted for any combination of $J_1$ and $J_2$ at all temperatures\cite{isakov}.

 In order to take into account any finite size scaling effects the Binder cumulant\cite{binder} was calculated for several values of $\frac{J_2}{J_1}$. The Binder cumulant is defined as
\begin{equation}
U_L = 1 - \frac{\langle m^4 \rangle}{3 \langle m^2 \rangle^2}\label{binderEq}
\end{equation}
where $m$ is the magnetization of the lattice. For the ferrimagnetic transition we have defined $\vec{m}=\sum_{j=1}^N{\vec{s} _j^{Cr}} - \sum_{k=1}^N{\vec{s} _k^{Fe}}$ which clearly demonstrates that in the region featuring two transitions, the high temperature transition indicates the onset of ferrimagnetic order. The cumulant has a critical temperature that is independent of the size of the lattice and is defined by the intersection point of the data acquired from simulations using different lattice sizes. The critical temperature obtained from the Binder cumulant is important as it defines a size invariant transition temperature for the lattice. This allows us to compare the transition temperatures extracted from the $L = 6$ heat capacities with results that should hold to the thermodynamic limit. Fig. \ref{figure2} shows that the transition obtained from the Binder cumulant differs from the values obtained from the heat capacity measurement by only a small positive amount. This small difference indicates that the transition temperatures obtained from the heat capacity were not caused by finite size effects and are reasonably close to the values that would be obtained in the thermodynamic limit. The $L = 9$ (not shown) results are seen to lie closer to the Binder results, indicating that the Binder results are consistent with the thermodynamic limit of our results.

\section{Heat Capacity}

\begin{figure}
\includegraphics[scale=0.35]{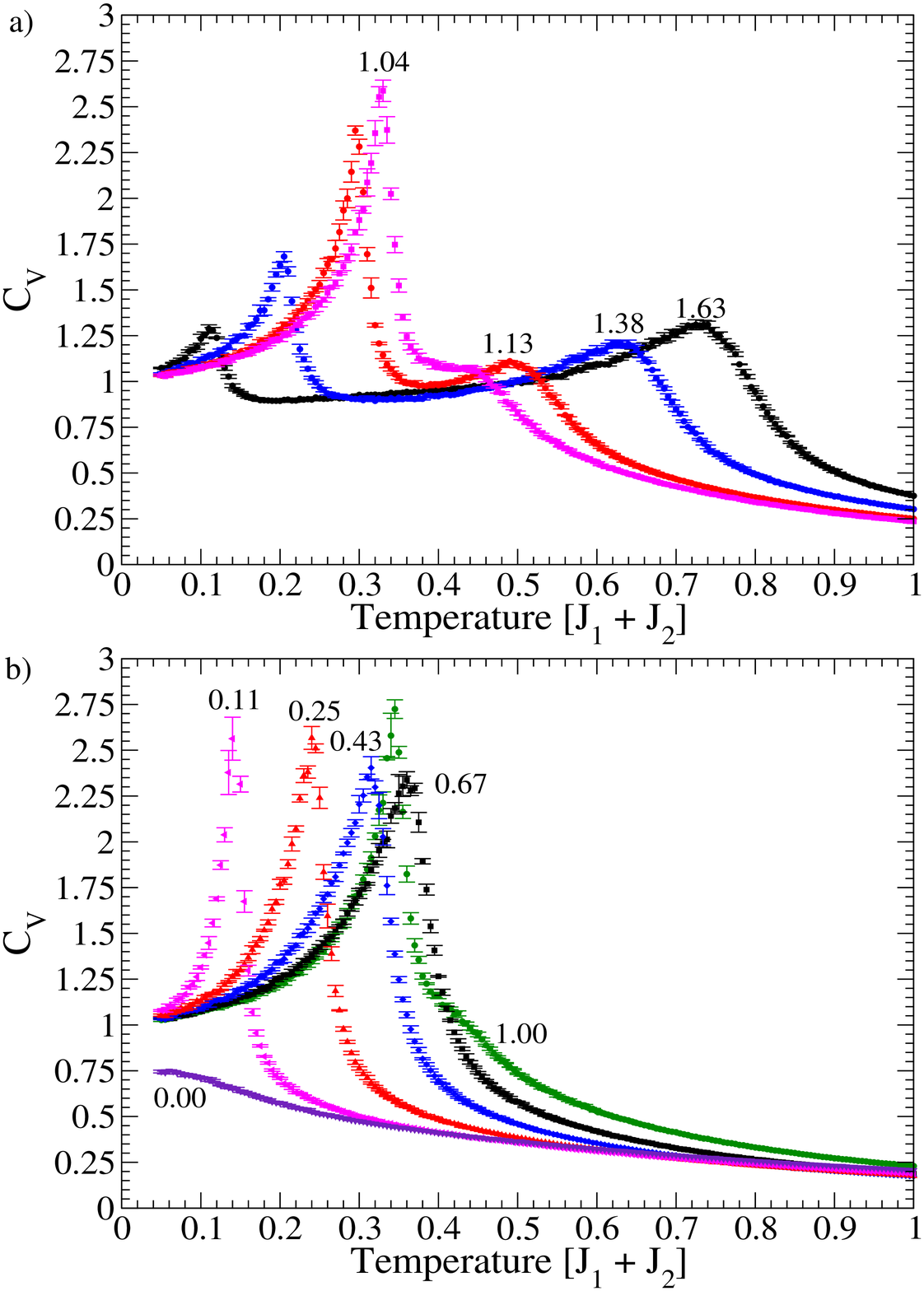}
\caption{(Color online) The heat capacity of the system at $L = 6$ for: a) $1 < \frac{J_2}{J_1} < 2$, and b) $\frac{J_1}{J_2} < 1$. The values of $\frac{J_2}{J_1}$ are labeled on each plot.\label{figure3}}
\end{figure}

 To determine the magnetic phase diagram of this model we have calculated the heat capacity as a function of temperature as a function of $J_1$ and $J_2$. From peaks in the heat capacity, we can extract the ordering temperature for various $J_1$ (with $J_1 + |J_2| = 1$). This allows us to write an approximate phase diagram delineated by the transition temperatures involved, as shown in Fig. \ref{figure2}. In the limit where $\frac{|J_2|}{J_1} > 2$ one has a ferri(ferro)magnetic form with spins in the Cr plane aligned and antialigned(aligned) to the spins in the Fe trimer. When $1 < \frac{|J_2|}{J_1} < 2$ (Fig. \ref{figure3}(a)) the Cr spins acquire both a component which has $120^\circ$ rotated spins and a component which tilts upward relative to the Cr plane, so the spin structure is canted. One appears to have two clear transitions. In the last region where $0 \le \frac{|J_2|}{J_1} < 1$ (Fig. \ref{figure3}(b)), all cases have a $120^\circ$ rotated spin structure in the Cr plane, and coplanar spins on the Fe trimers whose sum is opposite(parallel) to the spin sum of the nearest Cr sites.

 Representative plots of the heat capacity in units of the Boltzmann constant per atom as a function of temperature are shown in Fig. \ref{figure3}. As these are classical Heisenberg spins subject to the constraint of unit magnitude, one expects from the equipartition of energy theorem that at low temperatures the heat capacity will have two quadratic degrees of freedom per site, leading to $\frac{1}{2}k_B$ each, for a total $C_v=k_B$. At low temperature in real quantum materials this cannot be physical, however, it is hoped that the higher temperature classical phase transitions may remain relevant. In this sense we see that as we turn on the coupling between the Cr atoms (from $J_1=0$), the transition temperature decreases monotonically until the point $\frac{J_2}{J_1} = 2$ below which two transitions (see Fig. \ref{figure3}(a)) occur as a function of temperature. This coincides with the transition from a ferrimagnetic ground state to establishing a ground state with a component which rotates on each triangle of the lattice ($\phi\not=0$). At the high temperature transition, the ferrimagnetic orientation of the spins sets in, with the coplanar $120^\circ$ component attaining order at still lower temperatures. From the low temperature transition (see Fig. \ref{figure3}(b)) the transition temperature rises a little for the next two values of $\frac{J_2}{J_1} = 1$ and $\frac{J_2}{J_1} = \frac{2}{3}$ before decreasing monotonically towards zero temperature for the uncoupled layer case.

 The heat capacity of the kagome case ($\frac{J_2}{J_1} = 0$, Fig. \ref{figure3}(b)) does not approach $1$ but instead approaches $\frac{3}{4}$ as the temperature approaches zero as the curves for other values of $J_1$ do because of its normalization.  Energetically it is as if the Fe trimers have zero spin as they no longer have any contribution to the heat capacity of the lattice, yet we have included them in the normalization. No ordering transition is seen, although a weak peak is observed in the heat capacity at low temperature consistent with previous work done by Zhitomirsky {\it et al.}\cite{zhitomirsky} and others \cite{chalker}.

\section{Magnetic Susceptibility}

 To guide the development of a realistic model of FeCrAs, we have performed calculations of the static magnetic susceptibility as a function of the temperature, $\chi(T) = \frac{1}{N \, T} \displaystyle\sum_{i,j} \! < \! \vec{s_i}(T) \cdot \vec{s_j}(T) \! >$. It is anticipated that such studies of classical magnetism on a frustrated lattice may show qualitatively similar results to a quantum version of our model, allowing us to estimate realistic parameters for $J_1$ and $J_2$. The bulk of our analysis to date has been carried out with $J_2 \geq 0$. The magnetic susceptibility for $\frac{J_2}{J_1} < 0$ has been found to follow similar trends to that of the $\frac{J_2}{J_1} > 0$ case with the main difference being a reduction in the magnitude of the susceptibility.

\begin{figure}
\includegraphics[scale=0.34]{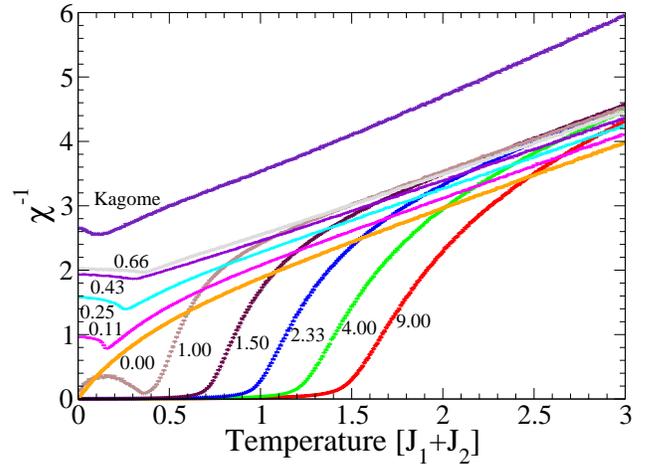}
\caption{(Color online) The inverse magnetic susceptibility as a function of temperature. For $\frac{J_2}{J_1} = 9.0$ to $1.5$, the inverse susceptibility quickly falls towards $0$. For $\frac{J_2}{J_1} = 1$ this quick descent is followed by a brief low temperature upturn. Qualitatively different physics is seen for $\frac{J_2}{J_1} \le \frac{2}{3}$, where one sees a high temperature linear dependence, followed by a low temperature downturn (likely due to the paramagnetic contributions of the Fe spin) before a lowest temperature increase toward a constant value. For comparison, the same calculation has been performed on the kagome lattice which agrees with results of Reimers {\it{et al.}}\cite{reimers}. The values of $\frac{J_2}{J_1}$ are labeled on the plot.\label{figure4}}
\end{figure}

 Since the $J_2 > 0$ magnetic susceptibility appears to more closely resemble the experimental data on FeCrAs\cite{wu} we will focus on that case here. In the limit that $\frac{J_2}{J_1}$ goes to zero the inverse magnetic susceptibility does not go to the kagome result as it contains a paramagnetic contribution due to the uncoupled Fe spins which reduces the inverse magnetic susceptibility to zero instead of reaching a constant, non-zero value as in the strictly distorted kagome planes case (see Fig. \ref{figure4}). The precise location of this downturn depends on the relative magnitudes of the spins on the Cr sites and the Fe trimers, here taken to be of unit magnitude. For small values of the coupling constant ($0<\frac{J_2}{J_1}<<1$) between the Fe and Cr layers the inverse magnetic susceptibility has features reminiscent of a lattice of distorted kagome planes, exhibiting a weak upturn followed by a decrease at low temperature. For small values of the in-plane coupling constant ($\frac{J_2}{J_1}>>1$) the inverse magnetic susceptibility displays properties of both limits. The inverse magnetic susceptibility of these intermediate values reduces towards zero as in the cases with small coupling within the Cr planes ($\frac{J_2}{J_1}>>1$), prior to increasing in a soft peak at lower temperatures for $\frac{J_2}{J_1}=1$, before returning to zero.

\begin{figure}
\includegraphics[scale=0.33]{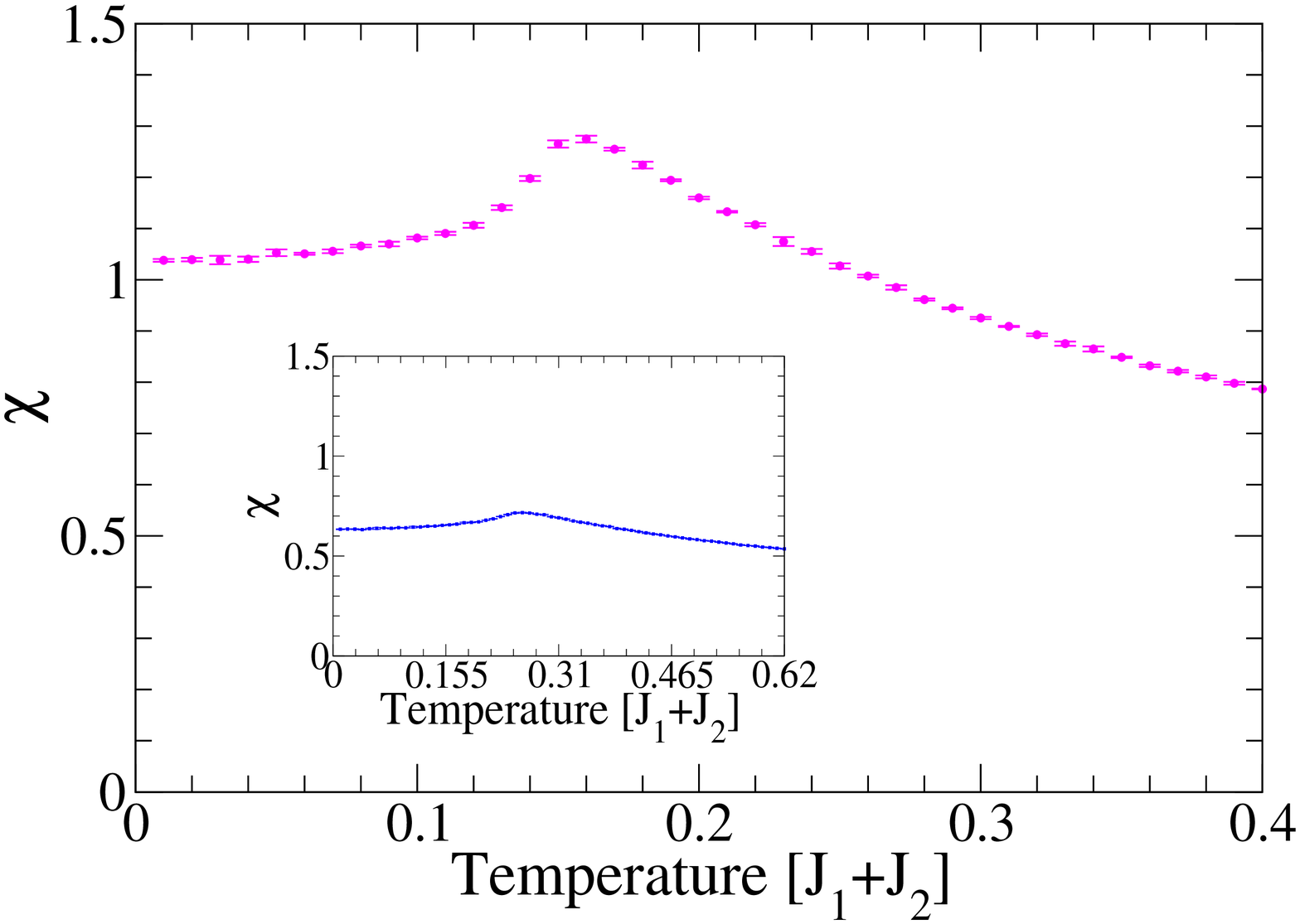}
\caption{(Color online) A plot of the magnetic susceptibility of the Monte Carlo results for $\frac{J_2}{J_1} = \frac{1}{9}$ and $\frac{J_2}{J_1} = \frac{1}{4}$ (inset).\label{figure5}}
\end{figure}

 To make contact with the experimental results of Wu {\it{et al.}}\cite{wu}, it is tempting to choose magnetic couplings $J_1$ and $J_2$ in such a way as to locate the maximum of the susceptibility roughly in the middle of our temperature scale as seen experimentally on a scale from $0$ to $300K$. While experimentally FeCrAs is seen to have an easy axis below the magnetic ordering temperature, which has not been included in our model, we see from Fig. \ref{figure5} that the qualitative features of the magnetic susceptibility seem to be reasonably well captured by $\frac{J_2}{J_1} \sim \frac{1}{9}$ when compared to Fig. 2 of Ref. \cite{wu}. The magnetic susceptibility of the Monte Carlo simulation appears qualitatively to be an average of the measured magnetic susceptibility along the two directions of the FeCrAs lattice. This comparison could be made quantitative by converting from dimensionless units to emu/(mole Oe) and including magnetic moments of larger relative magnitude on the Cr sites than the Fe trimers. While for frustrated magnets one often finds the inverse magnetic susceptibility to show a Curie-Weiss like straight line to extremely low temperatures, it is interesting to note that the presence of the Fe trimers appears to cause curvature near the ordering transition temperature. As such, it may not be necessary to subtract a temperature independent contribution to obtain the straightest possible $(\chi - \chi_0)^{-1}$ as carried out by Wu {\it{et al.}}\cite{wu}.

%\section{Outlook}
\section{Discussion and Summary}

 It appears that a classical spin model on the layered alternating distorted kagome and triangular lattice is able to capture magnetic order at a finite temperature with an ordering wavevector consistent with that observed experimentally in FeCrAs. 
Magnetic susceptibility simulations have features qualitatively consistent with measurements of Wu {\it{et al.}}\cite{wu} 
for $\frac{J_2}{J_1} \sim \frac{1}{9}$ if we assume unit spins on the Cr and Fe trimer sites. It is anticipated that a similar model with quantum spins on this lattice including higher orders of magnetic exchange interactions such as a $J_1$, $J_2$, and $J_3$ model where $J_3$ couples the Fe trimers may capture the low temperature heat capacity of this unusual non-Fermi liquid material.

It is also plausible that the low temperature phase seen in FeCrAs, featuring a Fermi liquid-like constant specific heat coefficient coinciding with insulating-like (power-law) electronic transport, indicates the presence of a U(1) spin liquid insulator with a finite spinon Fermi surface as first discussed by Podolsky {\it{et al.}} in the context of Na$_4$Ir$_3$O$_8$\cite{podolsky}. Within such a picture, charge neutral spinons lead to gapless excitations and the non-Fermi liquid behavior of transport may imply that FeCrAs is close to a quantum phase transition between metallic and insulating states. Such a possibility arising in the context of the model here presented of stacked distorted kagome and triangular planes will be presented elsewhere, along with a microscopic derivation of the model\cite{jrau}. We emphasize that the magnetic ordering obtained in this classical Heisenberg model is a robust feature independent of whether spinons or electrons are the proper description of the elementary magnetic units. Therefore when the moment size and/or orbital overlap with the Cr atoms decreases, the magnetic transition temperature should decrease, as the coupling between Fe and Cr is a crucial mechanism for the magnetic ordering. Such physics may be accessible in these materials under the application of uniaxial pressure.

In summary, we have studied a classical Heisenberg model on the hexagonal Fe$_2$P/ZrAlNi structure. We have focussed on the common situation where both the 3f and 3g sites hold magnetic atoms, with the 3g sites forming layers of distorted kagome planes bisected by triangular lattice planes. The 3f sites form trimers, small equilateral triangles, about each site of the triangular lattice in the intermediate planes and in our simple model have been treated as single sites of unit spin, as have each of the 3g sites. The model shows two distinct phase transitions depending on the magnitude of the ratio of the coupling constants $\frac{J_2}{J_1}$. For values of $\frac{|J_2|}{J_1}<1$ the 3g spins are $120^\circ$ rotated on the corners of the equilateral triangles, while the coplanar 3f spins align opposite(parallel) to the spin sum on the nearest 3g distorted hexagon. For $\frac{|J_2|}{J_1}>2$ the spins take on a ferri(ferro)magnetic form, ferromagnetic within each plane. For intermediate values sequential transitions are seen with a ferri(ferro)magnetic transition at high temperatures preceding a lower temperature $120^\circ$ rotation of some component of the canted spin. The ordering temperature has been extracted from the heat capacity plots for $0\le\frac{|J_2|}{J_1}\le\inf$ and a phase diagram has been presented, the validity of which in the thermodynamic limit has been checked by the calculation of a Binder cumulant. These results may be of relevance to FeCrAs with Cr atoms in the 3g and Fe in the 3f sites. We have calculated the magnetic susceptibility for a ratio of the coupling constants of $\frac{J_2}{J_1} = \frac{1}{9}$ was found to be qualitatively similar to the experimental results of Wu {\it et al.}\cite{wu}.

\vskip1pc
{\it{Acknowledgements:}} This work has been supported by NSERC (J.M.H. and H.Y.K.), BURC (J.M.H.) and CRC (H.Y.K.). We thank S. Julian for useful discussions related to experimental data.


\begin{thebibliography}{00}
\bibitem{sondhimoessner} C. Castelnovo, R. Moessner and S. Sondhi, Nature (London) {\bf{451}}, 42 (2008).
\bibitem{experimentalmonopoles} S. T. Bramwell, S. R. Giblin, S. Calder, R. Aldus, D. Prabhakaran and T. Fennell, Nature (London) {\bf{461}}, 956 (2009); T. Fennel, P. P. Deen, A. R. Wildes, K. Schmalzl, D. Prabhakaran, A. T. Boothroyd, R. J. Aldus, D. F. McMorrow and S. T. Bramwell, Science {\bf{326}}, 415 (2009).
\bibitem{us1} T. E. Redpath and J. M. Hopkinson, Phys. Rev. B {\bf{82}}, 014410 (2010).
\bibitem{us2} P. Carter, J. Hopkinson and M. Enjalran, APS March Meeting, 16 March 2009 (unpublished); T. E. Redpath, P. Carter, J. M. Hopkinson and M. Enjalran, unpublished.
\bibitem{Anderson87} P. W. Anderson, Science, {\bf{235}}, 1196 (1987).
\bibitem{TakagiPRL2007}  Y. Okamoto, M. Nohara, H. Aruga-Katori and H. Takagi, Phys. Rev. Lett. {\bf{99}}, 137207 (2007).
\bibitem{hopkinsonprl2007} J. M. Hopkinson, S. V. Isakov, H.-Y. Kee and Y. B. Kim, Phys. Rev. Lett. {\bf{99}} 037201 (2007).
\bibitem{lawler1} M. J. Lawler, H. Y. Kee, Y. B. Kim and A. Vishwanath, Phys. Rev. Lett. {\bf{100}}, 227201 (2008).
\bibitem{zhou} Y. Zhou, P. A. Lee, T.-K. Ng and F. C. Zhang, Phys. Rev. Lett. {\bf{101}}, 197201 (2008).
\bibitem{lawler2} M. J. Lawler, A. Paramekanti, Y. B. Kim and L. Balents, Phys. Rev. Lett. {\bf{101}}, 197202 (2008).
\bibitem{chen} G. Chen and L. Balents, Phys. Rev. B {\bf{78}} 094403 (2008).
\bibitem{bergholtz} E. J. Bergholtz, A. M. L{\"{a}}uchli and R. Moessner, Phys. Rev. Lett. {\bf{105}} 237202 (2010).
\bibitem{shores} M.P. Shores, E.A. Nytko, B.M. Bartlett and D.G. Nocera, J. Am. Chem. Soc. {\bf{127}} 13462 (2005).
\bibitem{helton} J. S. Helton, K. Matan, M. P. Shores, E. A. Nytko, B. M. Bartlett, Y. Yoshida, Y. Takano, A. Suslov, Y. Qiu, J.-H. Chung, D. G. Nocera and Y. S. Lee, Phys. Rev. Lett. {\bf{98}} 107204 (2007).
\bibitem{lee} S.-H. Lee, H. Kikuchi, Y. Qiu, B. Lake, Q. Huang, K. Habicht and K. Kiefer, Nature Mater. {\bf{6}} 853 (2007).
\bibitem{devries} M. A. de Vries, K. V. Kamenev, W. A. Kockelmann, J. Sanchez-Benitez and A. Harrison, Phys. Rev. Lett. {\bf{100}} 157205 (2008).
\bibitem{langasites} J. Robert, V. Simonet, B. Canals, R. Ballou, P. Bordet, P. Lejay and A. Stanault, Phys. Rev. Lett. {\bf{96}} 197205 (2006); {\bf{97}} 259901(E) (2006); V. Simonet, R. Ballou, J. Robert, B. Canals, F. Hippert, P. Bordet, P. Lejay, P. Fouquet, J. Ollivier and D. Braithwaite, Phys. Rev. Lett. {\bf{100}} 237204 (2008).
\bibitem{wu} W. Wu, A. McCollam, P.M.C. Rourke, D.G. Rancourt, Ian Swainson and S.R. Julian, EPL {\bf{85}} 17009 (2009).
\bibitem{swain} Ian P. Swainson, Wenlong Wu, Alix McCollam and Stephan R. Julian, Can. J. Phys. {\bf{88}} 10 (2010).
\bibitem{diep} See ex. {\it{Frustrated Spin Systems}}, edited by H. T. Diep (World Scientific, Singapore, 2004).
\bibitem{chalker} J.T. Chalker, P.C.W. Holdsworth and E.F. Shender, Phys. Rev. Lett. {\bf{68}} 855 (1992).
\bibitem{huse} D.A. Huse and A.D. Rutenberg, Phys. Rev. B {\bf{45}} 7536 (1992).
\bibitem{reimers} J.N. Reimers and A.J. Berlinsky, Phys. Rev. B {\bf{48}} 9539 (1993).
\bibitem{chubukov} A.V. Chubukov, Phys. Rev. Lett. {\bf{69}} 832 (1992).
\bibitem{harris} A.B. Harris, C. Kallin and A.J. Berlinsky, Phys. Rev. B {\bf{45}} 2899 (1992).
\bibitem{gondek} L. Gondek and A. Szytula, J. Alloys Comp. {\bf{442}} 111 (2007).
\bibitem{schmalfuss} D. Schmalfu\ss, J. Richter and D. Ihle, Phys. Rev. B {\bf{70}} 184412 (2004).
\bibitem{zelli} M. Zelli, K. Boesse and B.W. Southern, Phys. Rev. B {\bf{76}} 224407 (2007).
\bibitem{jrau} J. Rau and H.-Y. Kee, unpublished.
\bibitem{P321} Or in the trigonal $P321$ space group to which the langasites are believed to belong, which differs only in the c-axis position of its non-magnetic atoms and was originally thought to be the hexagonal Fe$_2$P strcuture (see ex. http://cst-www.nrl.navy.mil/lattice/struk/c22o.html).
\bibitem{baran} S. Baran, M. Hofmann, J. Leciejewicz, B. Penc, M. Slaski, A. Szytula and A. Zygmunt, J. Magn. Magn. Mater. {\bf{222}} 227 (2000).
\bibitem{baran2} S. Baran, M. Hofmann, J. Leciejewicz, B. Penc, M. Slaski, A. Szytula and A. Zygmunt, J. Alloys Compd. {\bf{281}} 92 (1998).
\bibitem{morosan} E. Morosan, S. L. Bud'ko and P. C. Canfield, Phys. Rev. B {\bf{71}} 014445 (2005).
\bibitem{bacmann} M. Bacmann, D. Fruchart, A. Koumina and P. Wolfers, Materials Science Forum, {\bf{443-444}} 379 (2004).
\bibitem{guerin} P. R. Gu{\'{e}}rin and M. Sergent, Acta Cryst. B {\bf{33}} 2820 (1977).
\bibitem{iwataki} T. Iwataki, H. Ohsato, K. Tanaka, H. Morikoshi, J. Sato and K. Kawasaki, J. Euro. Cer. Soc. {\bf{21}} 1409 (2001); P. Bordet, I. Gelard, K. Marty, A. Ibanez, J. Robert, V. Simonet, B. Canals, R. Ballou and P. Lejay, J. Phys.: Cond. Matt. {\bf{18}} 5147 (2006).
\bibitem{merminwagner} N.D. Mermin and H. Wagner, Phys. Rev {\bf 17} 1133 (1966).
\bibitem{isakov} as suggested by S.V. Isakov.
\bibitem{binder} K. Binder, Z. Phys. {\bf{B43}}, 119 (1981).
\bibitem{zhitomirsky} M.E. Zhitomirsky, Phys. Rev. B {\bf{78}} 094423 (2008).
\bibitem{podolsky} D. Podolsky, A. Paramekanti, Y.B. Kim and T. Senthil, Phys. Rev. Lett. {\bf{102}} 186401 (2009).
\end{thebibliography}
\end{document}